\documentclass[12pt]{article}
\usepackage{graphicx}
\usepackage{lineno}
\usepackage{hyperref}
\usepackage{caption}
\usepackage{subcaption}

\newcommand\pubnumber{ATL-PHYS-PROC-2024-006}
\newcommand\pubdate{\today}

\def\institute{Comenius University}
\def\authemail{\footnote{Contact: pavol.bartos@cern.ch \newline \newline Copyright 2023 CERN for the benefit of the ATLAS Collaboration. Reproduction of this article or parts of it is
allowed as specified in the CC-BY-4.0 license.}}

\def\Title#1{\begin{center} {\Large #1 } \end{center}}
\def\Author#1{\begin{center}{ \sc #1} \end{center}}
\def\Address#1{\begin{center}{ \it #1} \end{center}}

\newcommand\pubblock{\rightline{\begin{tabular}{l} \pubnumber\\
         \pubdate  \end{tabular}}}
\newenvironment{Abstract}{\begin{quotation}  }{\end{quotation}}
\newenvironment{Presented}{\begin{quotation} \begin{center} 
             PRESENTED AT\end{center}\bigskip 
      \begin{center}\begin{large}}{\end{large}\end{center} \end{quotation}}
\def\Acknowledgements{\bigskip  \bigskip \begin{center} \begin{large}
             \bf ACKNOWLEDGEMENTS \end{large}\end{center}}

\def\beq{\begin{equation}}
\def\eeq#1{\label{#1}\end{equation}}
\def\eeqn{\end{equation}}

\def\beqa{\begin{eqnarray}}
\def\eeqa#1{\label{#1}\end{eqnarray}}
\def\eeqan{\end{eqnarray}}

\let\bar=\overbar

\def\Dslash{\not{\hbox{\kern-4pt $D$}}}
\def\dslash{\not{\hbox{\kern-2pt $\del$}}}

\def\msb{{\bar{\ssstyle M \kern -1pt S}}}

\begin{document}
\begin{titlepage}
\pubblock

\vfill
\Title{Searches for FCNC and lepton flavour violating interactions of the top quark with the ATLAS detector}
\vfill
\Author{ Pavol Barto\v{s}\authemail{}, on behalf of the ATLAS Collaboration}
\Address{\institute}
\vfill
\begin{Abstract}
The LHC is a top quark factory and provides a unique opportunity to look for flavour changing neutral current or charged-lepton flavour violating interactions of the top quark. These processes are highly suppressed in the Standard Model and are beyond the experimental sensitivity, but can receive enhanced contributions in many extensions of the Standard Model. Results of searches for flavour changing neutral current $tqH$ vertex and charged-lepton flavour violating $\mu\tau{}tq$ vertex are presented. The searches find good agreement with the Standard Model expectation and derived exclusion bounds are improved very significantly.
\end{Abstract}
\vfill
\begin{Presented}
$16^\mathrm{th}$ International Workshop on Top Quark Physics\\
(Top2023), 24--29 September, 2023
\end{Presented}
\vfill

\end{titlepage}
\def\thefootnote{\fnsymbol{footnote}}
\setcounter{footnote}{0}

\section{Introduction}

Until 2012, the Higgs boson was the last undiscovered particle of the Standard Model (SM). The observation of the Higgs boson~\cite{atlasHiggs,cmsHiggs} at the Large Hadron Collider (LHC) by the ATLAS and CMS collaborations completed the SM. Nonetheless, the physics programme of the LHC experiments is still rich. The precision of the measurements of the Higgs boson and top quark properties is being improved, and searches for rare beyond the SM (BSM) processes are tightening the limits. A part of the programme are also searches for flavour changing neutral currents (FCNC) and charged-lepton flavour violation (cLFV) interactions involving Higgs boson and/or top quark. In this paper, we present the latest results of these searches provided by the ATLAS~\cite{atlas} experiment using 139 fb$^{-1}$ of proton-proton collisions at $\sqrt{s}=13\;$TeV. 

\section{Flavour changing neutral currents}
In the SM, the FCNC interactions are forbidden at the tree level and are very suppressed at the higher orders. For decay $t\rightarrow{}qX$, where $q$ stands for up-type quark (up or charm) and $X$ represents $H$ boson, $Z$ boson, gluon or photon, the predicted branching ratios are at the level of $10^{-16}$ -- $10^{-12}$~\cite{fcncSMpred}. However, if BSM scenarios are taken into account, these branching ratios are several orders of magnitude larger~\cite{fcncpred} and could be observed at the LHC. 

In the recent search for the FCNC~\cite{aFCNC_Hgammagamma}, the interactions of top quark, Higgs boson and up-type quark are studied in events with the Higgs boson decaying to two photons. The integrated luminosity of the data sample used in this analysis is increased by factor of about 4 with respect to previous ATLAS measurement~\cite{aFCNC1}. The following processes are investigated: associated production of single top quark and Higgs boson via FCNC vertex; and top-quark pair production with one of the top quark decaying according to the SM to $W$ boson and $b$ quark and another top quark decaying via the FCNC vertex to Higgs boson and up-type quark (up or charm). Events with two photons from Higgs boson decay and one $b$-tagged jet are divided into leptonic and hadronic channels, according to the decay products of $W$ boson. Further categorisation is based on e.g. number of jets, reconstructed invariant mass(es) of top quark(s) in order to distinguish between $pp \rightarrow tH$ and $pp\rightarrow t\bar{t}$ production. In the latter case, charm tagging is used to identify $t\rightarrow cH$ and $t\rightarrow uH$ processes. In addition, three different boosted decision trees are employed in different event categories to suppress the background and improve sensitivity of the analysis.
The branching ratio of decay $t\rightarrow c(u)H$ is determined in fit to data using a likelihood function. Two different likelihood constructions are used. In all but one event categories, standard unbinned extended likelihood is used, while the background shape is estimated from the distribution of the invariant mass of the two photons, $m_{\gamma\gamma}$, using MC simulations. In the one category, simple event counting is used due to lack of statistics in region of $m_{\gamma\gamma} \in [100,122]\cup [129,160]$ GeV. The data and SM predictions agree within uncertainties, i.e. no statistically significant excess due to FCNC coupling is observed. Using the CL$_\mathrm{s}$ method~\cite{CLs}, the observed (expected) $95\%$ confidence level (CL) upper limits on branching ratio are $4.3\times 10^{-4}$ ($4.7\times 10^{-4}$) for the $t\rightarrow cH$, and $3.8\times 10^{-4}$ ($3.9\times 10^{-4}$) for the $t\rightarrow uH$. The sensitivity of the analysis is limited by the statistical precision. The most relevant systematic uncertainties come from the non-resonant background estimation, photon energy resolution, $t\bar{t}$ cross-section, $H\rightarrow\gamma\gamma$ branching ratio, and parton shower description. Thanks to improved event reconstruction and event categorisation, the enhancement of the sensitivity of the analysis is about factor of 1.5 better than expected from the increase of the integrated luminosity. 

\begin{figure}[!h!tbp]
\centering
     \begin{subfigure}[b]{0.45\textwidth}
     \centering
     \includegraphics[width=\textwidth]{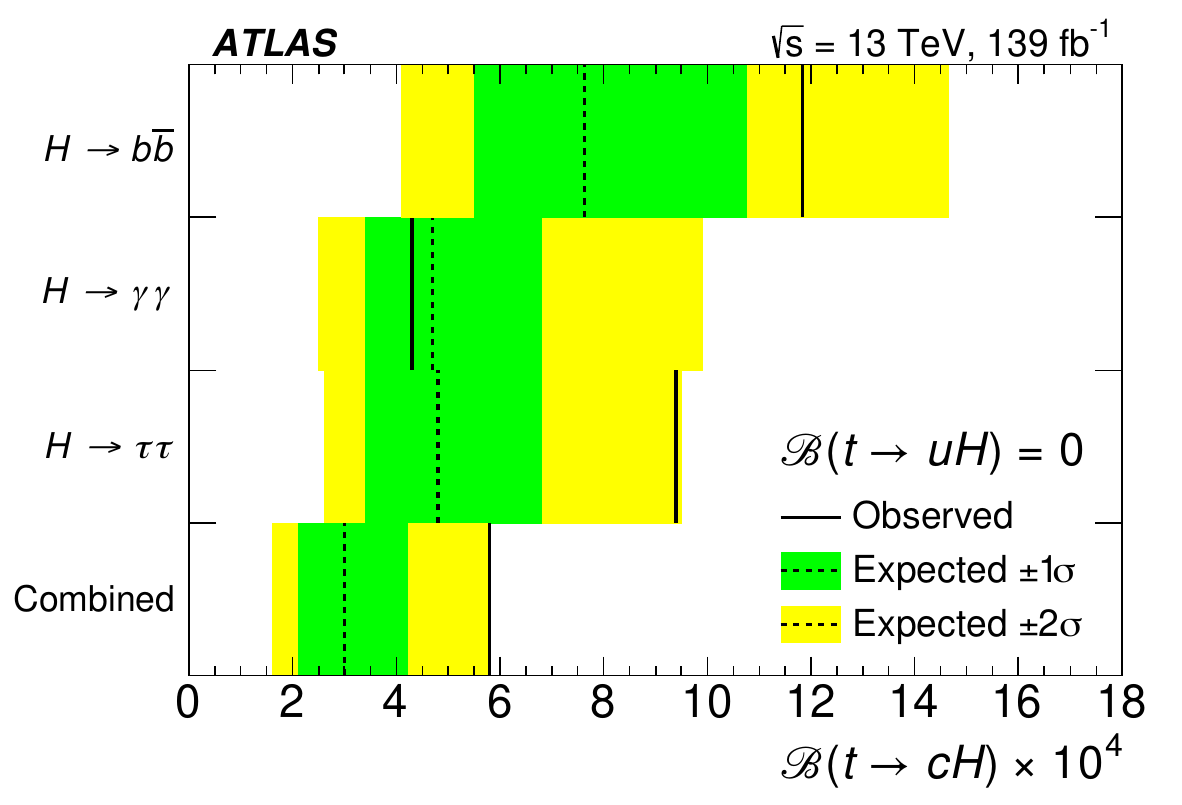}
     \caption{}
     \end{subfigure}
     \hfill
     \begin{subfigure}[b]{0.45\textwidth}
     \centering
     \includegraphics[width=\textwidth]{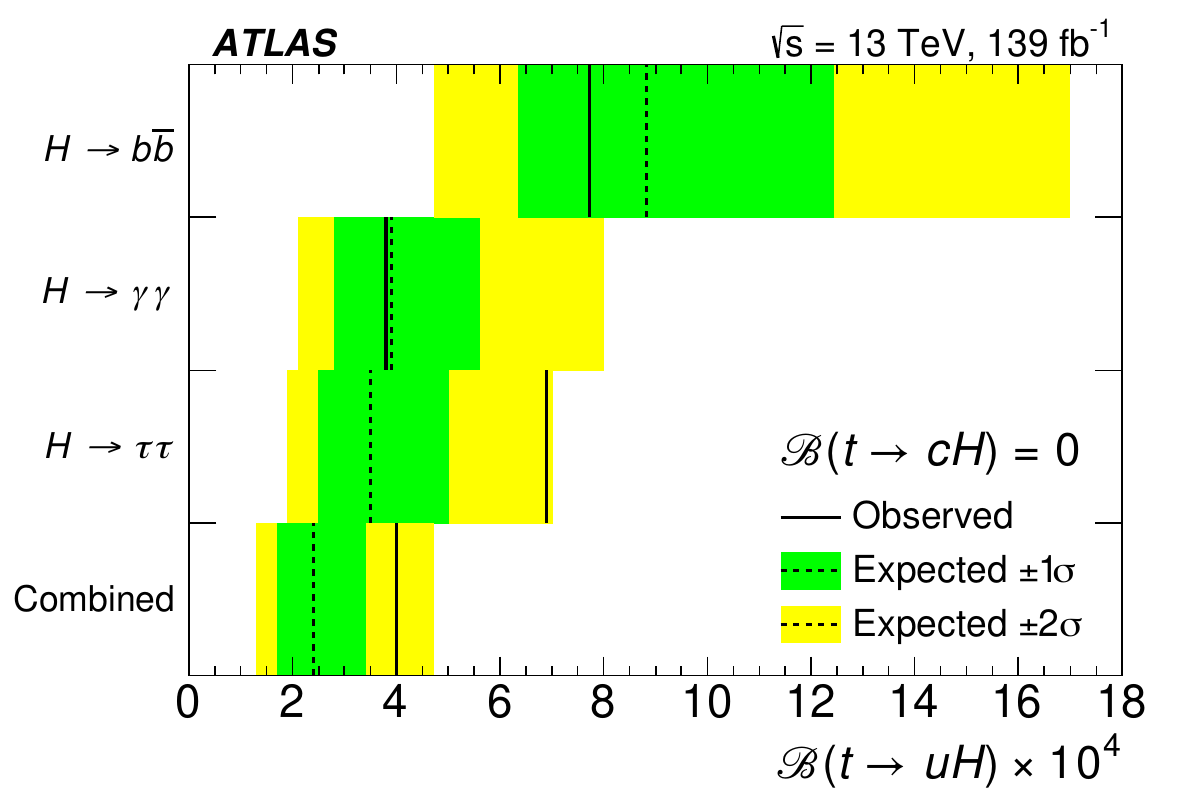}
     \caption{}
     \end{subfigure}
\caption{$95\%$ CL upper limits on (a) $\cal B$$( t \rightarrow cH)$ assuming $\cal{B}$$(t \rightarrow uH) = 0$ and (b) $\cal{B}$$(t \rightarrow uH)$ assuming $\cal{B}$$(t \rightarrow cH) = 0$ for the individual searches and their combination.}
\label{fig:tqHsummary}
\end{figure}

The recent $tqH$ analysis with $H\rightarrow \gamma\gamma$~\cite{aFCNC_Hgammagamma} is combined with the previous results from $tqH$ with $H\rightarrow b\bar{b}$~\cite{Htobb} and $H \rightarrow \tau\tau$~\cite{Htotautau}. The summary of the upper limits on the branching ratios from the individual measurements and the combination is shown in Figure~\ref{fig:tqHsummary}. As the individual measurements have different dominant systematic uncertainties and searches using the $H\rightarrow \gamma\gamma$ and $H \rightarrow \tau\tau$ decays are limited by statistical uncertainty, the combined results have low sensitivity to assumed correlations. The observed limits on the branching ratios from the combination are translated to the upper limits on the corresponding Wilson coefficients in the SM effective field theory: $C_{c\phi}=1.07$, $C_{u\phi}=0.88$~\cite{aFCNC_Hgammagamma}.

The $95\%$ confidence level observed limits on the branching ratios of the top quark decays via FCNC to up-type quark and a neutral boson provided by ATLAS are summarised  Ref.~\cite{fcnc_all}.

\section{Charged-lepton flavour violation}
Unless the neutrino oscillations were observed, the flavour of the charged and neutral leptons was expected to be conserved in the SM. Assuming higher order corrections in the SM, production rates of cLFV processes are far beyond current experimental sensitivity~\cite{cLFVpred}. In extensions of the SM, non-zero neutrino mass and mixing is provided together with local conservation of charged-lepton flavour. Observation of cLFV process would be a sign of new physics.

\begin{figure}[!h!tbp]
\centering
     \begin{subfigure}[b]{0.45\textwidth}
     \centering
     \includegraphics[width=\textwidth]{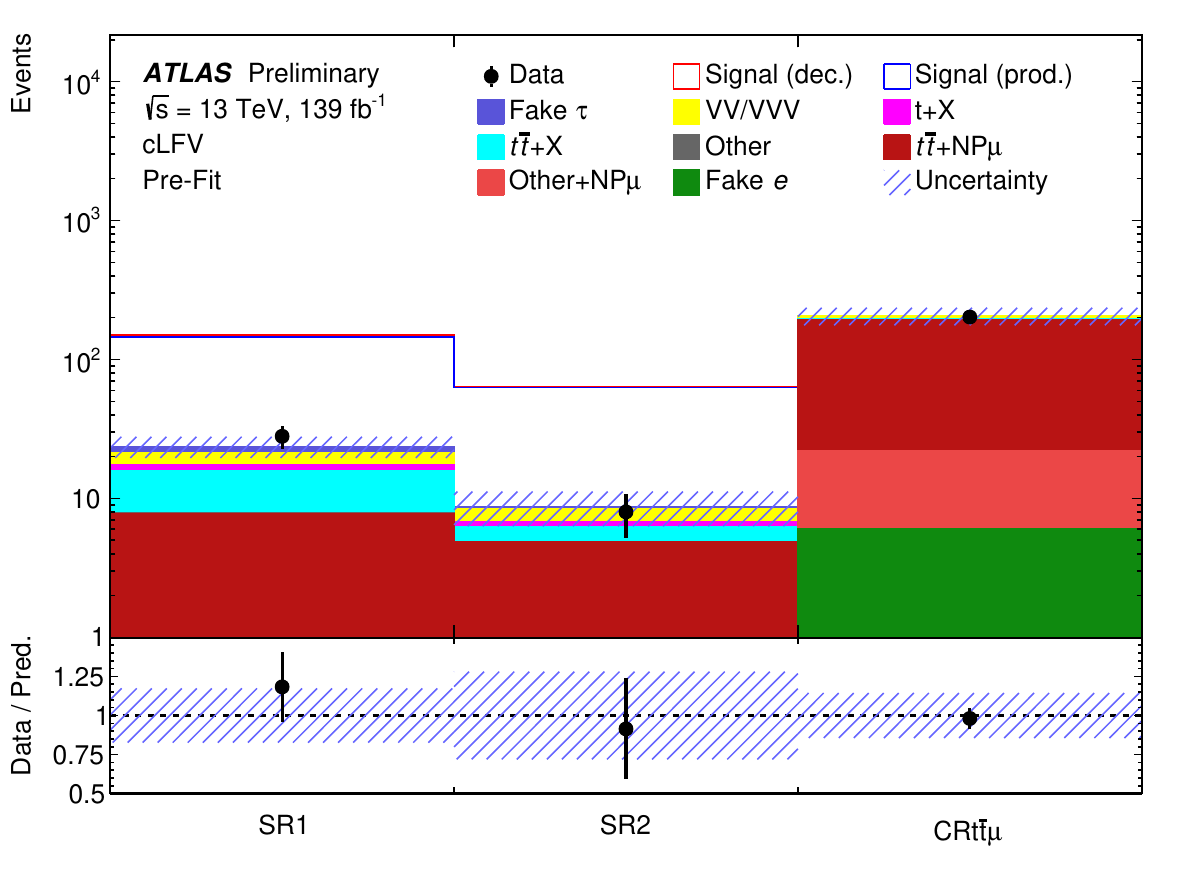}
     \caption{}
     \end{subfigure}
     \hfill
     \begin{subfigure}[b]{0.45\textwidth}
     \centering
     \includegraphics[width=\textwidth]{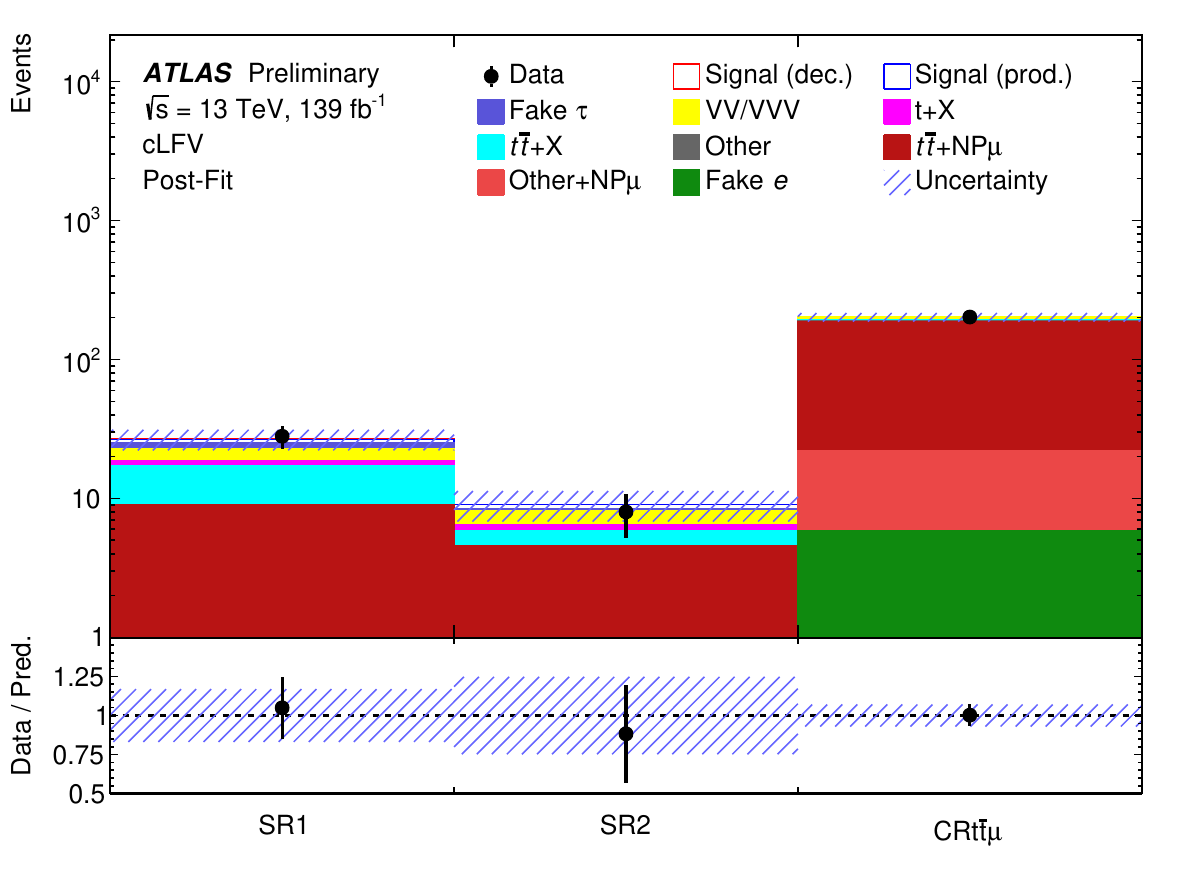}
     \caption{}
     \end{subfigure}
\caption{Observed event yields compared to (a) pre-fit and (b) post-fit expectations from Monte Carlo simulations. SR1 (SR2) represents processes with $t\rightarrow \ell^{\pm}\ell'^{\mp}q_k$ decay ($gq_k \rightarrow t\ell^{\pm}\ell'^{\mp}$ production), while CR$t\bar{t}\mu$ corresponds to the control region for dominant background.}
\label{fig:cLFV}
\end{figure}

In the recent analysis~\cite{cLFV}, $pp\rightarrow t\bar{t}$ (with $t\rightarrow \ell^{\pm}\ell'^{\mp}q_k$ decay) and $gq_k \rightarrow t\ell^{\pm}\ell'^{\mp}$ production 
processes are investigated, where $q_k$ stands for up-type quark (up or charm) and opposite signed leptons $\ell^{\pm}\ell'^{\mp}$ represent pair $\tau\mu$ or $\mu\tau$. In the event selection, presence of the hadronically decaying $\tau$, at least one jet and exactly one $b$-jet is required. In addition, there have to be two same-sign muons in the event - one from the semileptonic decay of top quark, another from cLFV vertex. Selected events are divided into two signal regions, one for each production process. A control region is defined for the dominant background process, which is dilepton decay channel of $t\bar{t}$ production with additional non-prompt $\mu$ from heavy-flavour decay inside a jet. The event yields in the two signal regions and background control region are shown in Figure~\ref{fig:cLFV}. The normalisation factors for signal and dominant backgroud contributions are obtained by simultaneous profile-likelihood fit. The measured data and SM predictions agree within the uncertainties and no significant cLFV contributions are observed. The results are interpreted 
as constraints on the Wilson coefficients related to top quark 2-quark-2-lepton operators involving $\mu$ and $\tau$ in the effective field theory, which were highly unconstrained prior this measurement. The obtained upper limit on $\cal B$$(t\rightarrow \mu\tau q)$ is $11\times 10^{-7}$ at $95\%$ CL. 

\section*{Conclusions}
The recent search for FCNC in the $tqH$ processes with $H$ decaying to two photons provide improved limits on the $\cal B$$(t\rightarrow u(c)H)$. After combining these results with the results from searches using $tqH$ events with the $H\rightarrow b\bar{b}$ and $H\rightarrow \tau\tau$ decays, more stringent limits on the Wilson coefficients are obtained. The recent search for cLFV processes presents constraints on the Wilson coefficients improved by factor of $8 - 51$ (depending on the coefficient). The observed upper limit on the $\cal B$$(t\rightarrow \mu\tau q)$ complements the CMS searches for cLFV in the $e\mu qt$ interactions.

\Acknowledgements
I am grateful to the Slovak Ministry of Education, Science, Research and Sport for the support and to my colleagues M. Dubovsk\'{y}, B. Eckerov\'{a}, and S. Tok\'{a}r for their valuable comments.

\bibliography{eprint}{}
\bibliographystyle{unsrt}
 
\end{document}